\documentclass[pop,fleqn,finallayout]{w-art}
\usepackage{amssymb}

\newcommand{\CC}{{\mathcal C}}
\newcommand{\HH}{{\mathcal H}}

\begin{document}
\Volume{49}
\Issue{10--11}
\copyrightinfo{2001}{WILEY-VCH Verlag GmbH \& Co. KGaA, Weinheim}
\Month{10}
\Year{2001}
\pagespan{1019}{1025}

\title{Teleportation with partially entangled states}
\author[Z. Kurucz]{Z. Kurucz\inst{1}}
\address[\inst{1}]{%
  Department of Nonlinear and Quantum Optics, \\
  Research Institute for Solid State Physics and Optics, \\ 
  Hungarian Academy of Sciences, P.O. Box 49, H-1525 Budapest, Hungary}

\author[M. Koniorczyk]{M. Koniorczyk\inst{1,2}}
\address[\inst{2}]{%
  Institute of Physics, University of P\'ecs, Ifj\'us\'ag \'ut 6.
  H-7624 P\'ecs, Hungary} 

\author[J. Janszky]{J. Janszky\inst{1}}

\begin{abstract}
  The relations of antilinear maps, bipartite states and quantum channels is
  summarized. Antilinear maps are applied to describe bipartite states and
  entanglement. Teleportation is treated in this general formalism with an
  emphasis on conditional schemes applying partially entangled pure states.  It
  is shown that in such schemes the entangled state shared by the parties, and
  those measured by the sender should ``match'' each other.
\end{abstract}

\subjclass[pacs]{03.67.-a, 03.65.Ud, 03.67.Hk}
\maketitle

\section{Introduction}
\label{sec:intro}

A fundamental part of quantum mechanics is the description of the evolution of
quantum states. In the lack of measurements, the evolution of the quantum
state of closed systems can be described by unitary operators.  Measurements
are described by projections onto the eigenstates of the measured quantity. In
more general approach to the question of quantum state evolution, the system
in argument is considered as a part of a larger subsystem. This approach gives
rise to concepts of general quantum channels and generalized (POVM)
measurements.

In the context of quantum information and communication, quantum entanglement
is a central issue. General quantum kinematics of multipartite systems harbors
lots of secrets still. Quantum teleportation\cite{prl70_1895} is probably the
most frequently quoted application of entanglement, and its experimental
feasibility\cite{nature390_575,prl80_1121,science282_706} has further
increased its relevance.

The methods developed for representations of quantum channels can be used
successfully in the description of entanglement and teleportation. The results
in this paper are motivated by these kind of methods. Section
\ref{sec:formalism} summarizes some facts concerning the relations between
quantum states of bipartite systems, channels, antilinear and antiunitary
maps. Some of these have already found application in quantum information
theory, but those use a fixed antilinear map, which is related to a specific
maximally entangled state. We also present another possibility, namely we
consider different antilinear maps, which is an alternative description of all
pure bipartite states. This approach provides us with a very convenient
description of quantum teleportation, including all schemes applying a pure
(but not necessarily maximally entangled) resource.

Popescu~\cite{prl72_797} pointed out that teleportation is also possible using
mixed states but with a fidelity less than 1.  In a recent paper of
Banaszek~\cite{pra62_024301} a protocol using a partially entangled state was
optimized for average fidelity.  Horodecki {\it et al}~\cite{pra60_1888}
presented a formula for the fidelity of such imperfect teleportation schemes.

In conditional schemes on the other hand, fidelity can be one but at the cost
that the process sometimes fails, so the probability of successfulness (also
called efficiency) is less than one.  This is the case in conclusive
teleportation~\cite{9608005,9906039}, where partially entangled or even mixed
states and generalized measurements (POVM) are considered.  Using nonunitary
transformation at Bob's side also makes the process
probabilistic~\cite{pra61_034301}.

The main part of our paper is based on the formalism summarized in Section
\ref{sec:formalism}. Our description is completely independent of the
dimensions of the Hilbert-spaces involved, and we do not even need to fix a
basis. In Section \ref{sec:teleport} we give a general condition for
conditional teleportation in terms of the applicable entangled states and
joint measurements.  In Section \ref{sec:matching} we show that partially
entangled states are capable of conditional teleportation with fidelity one,
but only if the outcome of the measurement performed by Alice and the state
shared by the parties ``match'' each other.  Section \ref{sec:concl}
summarizes our results.

\section{States, channels and antilinear maps}
\label{sec:formalism}

Consider a bipartite system with subsystems A and B. The subsystems are
described by the Hilbert-spaces ${\cal H}_A$ and ${\cal H}_B$, thus the pure
states of the system are in ${\cal H}_A\otimes {\cal H}_B$. Let $\dim {\cal
  H}_A=\dim {\cal H}_B=N$. Let $\{|i\rangle_A\}$ and
$\{|i\rangle_B\}\ (i=0,\ldots ,N-1)$ denote the computational bases on ${\cal
  H}_A$ and ${\cal H}_B$, and let
\begin{equation}
  \label{eq:maxent}
|\Psi^+\rangle_{AB}=\frac{1}{\sqrt{N}}\sum_{i=0}^{N-1}|i\rangle_A|i\rangle_B
\end{equation}
be a maximally entangled state of the system. All other maximally entangled
states of the system can be obtained from $|\Psi^+\rangle$ by local unitary
transformations. The set of density matrices of system A will be denoted by
${\cal S}_A$. A quantum channel $\$ _A $ is a completely positive,
trace-preserving, and hermiticity preserving 
${\cal S}_A \rightarrow {\cal S}_A$ map. 

Keeping $|\Psi^{(+)}\rangle$ in mind, we may introduce the relative state
representation~\cite{pra54_2614} of states and operators on ${\cal H}_A$.  Any
pure state $|\Psi\rangle_A\in {\cal H}_A$ can be described by an
(unnormalized) \emph{index state} $|\Psi^*\rangle_B\in {\cal H}_B$ so that
\begin{equation}
  \label{eq:relrep_state}
  |\Psi\rangle_A=\, _B\langle \Psi^*|\Psi^+\rangle_{AB}.
\end{equation}
The state in argument is obtained as a partial inner product of its index
state and the maximally entangled state $|\Psi^+\rangle$. The mapping
creating the index state from the original state,
\begin{equation}
  \label{eq:Lpsip}
  L_{|\Psi^+\rangle}: {\cal H}_A\rightarrow {\cal H}_B,\quad 
  L_{|\Psi^+\rangle}|\Psi\rangle_A = |\Psi ^*\rangle_B
\end{equation}
is antilinear, and in fact $\sqrt{N} L_{|\Psi^+\rangle}$ is antiunitary.
Indeed, expanding an arbitrary $|\Psi\rangle_A$ on the computational basis,
\begin{equation}
  \label{eq:Lpsip_bas}
  L_{|\Psi^+\rangle}|\Psi\rangle_A=L_{|\Psi^+\rangle}\sum_i C_i|i\rangle_A=
 \frac{1}{\sqrt{N}} \sum_i C_i^*|i\rangle_B,
\end{equation}
from which the above properties follow.

The introduction of $L$ via $|\Psi^+\rangle$ is also useful in describing
channels $\$ _A$. Let us have the compound system in the state
$|\Psi ^+\rangle_{AB}$, and send subsystem A through the channel $\$ _A$ while
doing nothing with subsystem B. The effect of the channel on any pure state
$|\Psi\rangle_A$ of system A is then obtained by the partial inner product
with the corresponding index state:
\begin{equation}
  \label{eq:Relrep_chann}
  \$_A\left(|\Psi\rangle_A\, _A\langle \Psi |\right)=
N\, _B\langle \Psi ^*|(\$ _A\otimes I_B)
\left (|\Psi^+\rangle _{AB} \,_{AB} \langle \Psi^+|\right)
|\Psi ^* \rangle_B,
\end{equation}
where $|\Psi^*\rangle_B=L_{|\Psi^+\rangle}|\Psi\rangle_A$, and $I_B$ stands
for the identity operator.  This is the so called relative
state representation of channels, which is widely used to describe them.
But even more can be stated~\cite{pra60_1888}. 
An affine isomorphism between the set of all $\$ _A$ channels on ${\cal S}_A$,
and the set of bipartite states 
$\varrho _{AB}\in {\cal S}_{{\cal H}_A\otimes{\cal H}_B}$ 
with maximally mixed partial trace, i.e. with the property
\begin{equation}
    \label{eq:parctrac}
    \mathop{\mbox{tr}}\nolimits _A \varrho_{AB}= \frac{1}{N} I_B,
\end{equation}
can be found similarly to Eq.~(\ref{eq:Relrep_chann}).
The bipartite state corresponding to a channel can be obtained from
$|\Psi^+\rangle$ by applying the channel on system A and doing nothing with
system B:
\begin{equation}
  \label{eq:Relrep_izom}
\varrho _{AB}=(\$ _A\otimes I_B)
\left( |\Psi^+\rangle_{AB} \, _{AB}\langle \Psi^+| \right). 
\end{equation}
As $|\Psi^+\rangle_{AB}$ has a maximally mixed partial trace, and this
property cannot be changed by local operations, (\ref{eq:parctrac}) will also
hold for the $\varrho _{AB}$ obtained in Eq.~(\ref{eq:Relrep_izom}).  The
isomorphism has been found by the Horodecki {\it et al.}~\cite{pra60_1888},
who have discussed it in detail, and have used it for the description of
teleportation channels.

So far we have considered the antiunitary map $\sqrt{N}L_{|\Psi^+\rangle}$
arising from the maximally entangled state $|\Psi^+\rangle$. This is a useful
tool in the description of channels and states. Let us follow the reverse way
now. We may use the set of antilinear ${\cal H}_A\rightarrow {\cal H}_B $ maps
in order to describe pure states in ${\cal H}_A\otimes {\cal H}_B$. As
relative state representation is also based on $L$, changing this map can give
rise to different relative state representations.

Consider a bipartite pure state $|\Phi\rangle_{AB}\in {\cal H}_A\otimes
{\cal H}_B $. We may write it on the computational basis as
$|\Phi\rangle_{AB}=\sum_{ij} C_{ij} |i\rangle_A \otimes |j\rangle_B$.  We
define the ${\cal H}_A\rightarrow {\cal H}_B $ antilinear operator
$L_{|\Phi\rangle}$ such that $L_{|\Phi\rangle}|i\rangle_A= \sum_j C_{ij}
|j\rangle_{B}$.  Thus we can write
\begin{equation}
  \label{eq:statdef}
  |\Phi\rangle_{AB}=\sum_i|i\rangle_A\otimes
   (L_{|\Phi\rangle}|i\rangle_A). 
\end{equation}
For any bipartite pure state $|\Phi\rangle_{AB}\in {\cal H}_A\otimes {\cal
  H}_B $, there uniquely exists an antilinear operator $L_{|\Phi\rangle}$
defined this way.

Because of the antilinearity of $L_{|\Phi\rangle}$, (\ref{eq:statdef}) is
independent of the actual computational basis chosen on ${\cal H}_A$. Let
$\CC_{AB}$ denote the set of bounded antilinear operators $L \colon
{\cal H}_A \to {\cal H}_B$ which have finite norm (that is,
$\mathop{\mbox{tr}}(L^\dag L) < \infty$, where the adjoint of $L$ is defined
by the relation $\langle f|Le \rangle = \langle L^\dag f|e \rangle^\ast$):
\begin{equation}
  \CC_{AB} = \big\{ L\colon {\cal H}_A \to {\cal H}_B\;
  \mbox{bound antilinear}\, 
  \big| \mathop{\mbox{tr}}(L^\dag L) < \infty \big\}.
\label{eq:CBC}\end{equation}
$\CC_{AB}$ forms a Hilbert space (the scalar product is $(L,L') =
\mathop{\mbox{tr}}(L'^\dag L)$, which is conjugate linear in the first
argument).  It is shown in Ref.~\cite{jmp41_638} that (\ref{eq:statdef})
establishes a unitary isomorphism between $\CC_{AB}$ and ${\cal H}_A\otimes
{\cal H}_B$ in a natural way.  Every pure bipartite state $|\Phi\rangle_{AB}$
can uniquely be described by an antilinear operator
$L_{|\Phi\rangle}\in \CC_{AB}$ such that
$\mathop{\mbox{tr}}(L_{|\Phi\rangle}^\dag L_{|\Phi\rangle})=1$.
Conversely, every such $L$ describes a pure bipartite state.

Now let us characterize maximally entangled states and possible
relative state representations. By maximally entangled state we mean a
pure bipartite state with maximally mixed partial trace
(\ref{eq:parctrac}). The partial traces of bipartite states over
systems $A$ and $B$ are $LL^\dag$ and $L^\dag L$ respectively.  Thus
the state (\ref{eq:statdef}) is maximally entangled if and only if
$LL^\dag=N^{-1} I_B$ and $L^\dag L=N^{-1} I_A$.  This is equivalent to
that $\sqrt{N}L$ is antiunitary.  On the other hand, the operator
$L_{|\Phi\rangle}$ gives rise to a relative state representation if
and only if $\{ L_{|\Phi\rangle}|i\rangle \}_{i=0,\ldots ,N-1}$ forms
an orthogonal basis on ${\cal H}_B$, which means, that
$\sqrt{N}L_{|\Phi\rangle}$ is antiunitary.  Relative state
representations can be defined via maximally entangled states.

\section{Probabilistic teleportation with partially entangled states}
\label{sec:teleport}

In this section we apply the antilinear description of bipartite states
introduced in Section \ref{sec:formalism} for quantum teleportation.  Suppose
that system A prepared in the unknown state $|\Phi\rangle_A$ is to be
teleported, and systems B and C shared by the parties (Alice and Bob) are in a
partially entangled state $|\sigma\rangle_{BC}$. We will call this
\emph{shared} state in what follows. The shared state is described by the
antilinear map $L_{|\sigma\rangle}$.  Systems A and B are located at Alice who
performs a joint projective measurement on them.  Suppose that its outcome
corresponds to the projection onto the state $|\sigma_q\rangle_{AB}$. In the
followings, we will regard only this outcome, thus our teleportation scheme
will be probabilistic, conditional one.

To have common computational bases in the description of the shared state and
the state the measurement project onto, we expand $|\sigma_q\rangle$ in the
following way:
\begin{equation}
   |\sigma_q\rangle = \sum_i (L_q |i\rangle) \otimes |i\rangle,
\label{eq:measdef}
\end{equation}
where $L_q \in \CC_{BA}$.  One can characterize the states corresponding to
nondegenerate measurement outcomes by bounded antilinear operators $L_q \colon
\HH_B \to \HH_A$ such that ${\mathrm{tr}}(L_q^\dag L_q)=1$.  Note that $L_q$ is
unique disregarding a unit complex phase factor.

The projection resolution of every joint observable of $A$ and $B$, which has
nondegenerate eigenvalues, is (up to phase factors) uniquely described by an
orthonormal basis $L_q$ in $\CC_{BA}$.  Those measurements whose nondegenerate
outcomes are represented by projections onto mutually orthogonal maximally
entangled states, are called measurements of Bell type~\cite{pla272_32}.
Every Bell measurement can be described by an orthonormal basis $L_q$ in
$\CC_{BA}$ such that $(\dim\HH_A)^{1/2} L_q$ is antiunitary for every $q$.

Now we will calculate the teleportation channel. By this we mean a function
$f_q \colon \HH_A \rightarrowtail \HH_C$ that relates the input state, and the
state of system C after the measurement. Note that, although the reversing
unitary transformation is usually also included in the definition of
teleportation channel, our terminology is more convenient here, as we
investigate linearity and reversibility. At the beginning, the three systems
are in the state $|\Phi\rangle_A \otimes |\sigma\rangle_{BC}$.  The
probability of the outcome $q$ under consideration is given by
\begin{eqnarray}
  p_q(|\Phi\rangle_A) &=& \left\| \strut [(|\sigma_q\rangle_{AB}
  \,_{AB}\langle\sigma_q|) \otimes I_C)]  (|\Phi\rangle_A \otimes
  |\sigma\rangle_{BC}) \right\| ^2 = \left\| {\sum\limits_i}
  \big( {}_A \langle \Phi | L_q | i \rangle_B^\ast \big)  
  L |i\rangle_B \right\|^2 
\nonumber\\
  &=& \left\| {\sum\limits_i} L( |i\rangle_B\,_B \langle i |  
  L_q^\dag |\Phi \rangle_A) \right\|^2 = \left\| \strut LL_q^\dag
  |\Phi\rangle _A  \right\|^2.
\label{eq:p_q}\end{eqnarray}
On condition that the measurement yields the outcome $q$, the state of system
$C$ can be written as
\begin{equation}
  \frac1{\sqrt{p_q(|\Phi\rangle_A)}} \sum_i \big( {}_{AB}\langle
  \sigma_q |   \Phi  \rangle_A |i\rangle_B \big)  
  L | e_i\rangle_B = \frac1{\sqrt{p_q(|\Phi\rangle_A)}} LL_q^\dag
  |\Phi\rangle_A. 
\end{equation}
The teleportation channel for the outcome~$q$ is
\begin{equation}
  f_q \colon \HH_A \rightarrowtail \HH_C, \quad 
  f_q(|\Phi\rangle_A) = \frac{LL_q^\dag |\Phi\rangle_A}{\left\| 
  LL_q^\dag |\Phi\rangle_A \right\|}. 
\label{eq:f_q}\end{equation}
If the input state is given by the density operator
$\rho_{\mathrm{in}}$ then the probability of the outcome $q$ is
\begin{equation}
  p_q(\rho_{\mathrm{in}}) = {\mathrm{tr}}_A\left( L_q L^\dag L L_q^\dag
  \rho_{\mathrm{in}} \right)
\label{eq:p_q(rho)}\end{equation}
and the output state is
\begin{equation}
  \rho_{\mathrm{out}} = \frac 
  {L L_q^\dag \rho_{\mathrm{in}} L_q L^\dag}
  {{\mathrm{tr}}_A\big( L_q L^\dag L L_q^\dag
  \rho_{\mathrm{in}} \big)}.
\label{eq:rhoout}\end{equation}

We have defined a special quantum operation based on the teleportation
scheme of Ref.~\cite{prl70_1895}.  One can obtain from
(\ref{eq:p_q(rho)}) that this operation is a generalized (POVM)
measurement of the input state and the positive operator representing
it is $L_q L^\dag L L_q^\dag$.

The channel $f_q$ has to be reversible, so that we can obtain a teleported
state identical to the original input state. We call the channel $f_q$
reversible, if it is injective, that is, for different input state
$|\Phi\rangle_A$ ($\||\Phi\rangle_A\|=1$) the corresponding output state
$f_q(|\Phi\rangle_A)$ is different. We remark, that the reversibility of
teleportation channels has also been investigated in Ref.~\cite{pra55_2547}.
We adopt a more general definition here. Reversibility means that every input
state can be recovered (theoretically) from the output state. One can easily
verify that this condition is equivalent to that the linear operator
$LL_q^\dag \colon \HH_A \to \HH_C$ is injective.

It may be the case, however, that the channel $f_q$ is not linear.
This way, the input state can be recovered from the output only using some
sophisticated nonlinear transformations, which may not be realistic.
Therefore, it is a natural requirement for the channel to be linear.

We show that if the teleportation channel is reversible, then its linearity is
equivalent to that the probability (\ref{eq:p_q}) of the outcome~$q$ is
independent of the input state $|\Phi\rangle_A$.  Suppose that
$|\Phi\rangle_1$ and $|\Phi\rangle_2$ are linearly independent, and let
$(\alpha_1|\Phi\rangle_1 + \alpha_2|\Phi\rangle_2)$ be such that
$\|\alpha_1|\Phi\rangle_1 + \alpha_2|\Phi\rangle_2\|=1$.  From the linearity
condition $f_q(\alpha_1 |\Phi\rangle_1 + \alpha_2 |\Phi\rangle_2) = \alpha_1
f_q(|\Phi\rangle_1) + \alpha_2 f_q(|\Phi\rangle_2)$, one can obtain:
\begin{multline}
  \alpha_1 \left( \frac1{\big\| LL_q^\dag (\alpha_1|\Phi\rangle_1 +
  \alpha_2|\Phi\rangle_2) \big\|} - \frac1{\big\| LL_q^\dag |\Phi\rangle_1
  \big\|} \right) LL_q^\dag |\Phi\rangle_1 
\\
  + \alpha_2 \left( \frac1{\big\|
  LL_q^\dag (\alpha_1|\Phi\rangle_1 + \alpha_2|\Phi\rangle_2) \big\|} -
  \frac1{\big\| LL_q^\dag |\Phi\rangle_2 \big\|} \right) LL_q^\dag
  |\Phi\rangle_2  =0.
\label{eq:linfq2}
\end{multline}
Since $f_q$ is injective, $LL_q^\dag |\Phi\rangle_1$ and $LL_q^\dag
|\Phi\rangle_2$ are also linearly independent.  Then (\ref{eq:linfq2})
implies that their coefficients are zero, that is, the probability
(\ref{eq:p_q}) of the outcome~$q$ is independent of the input state
$|\Phi\rangle_A$.  Reversely, if (\ref{eq:p_q}) is independent of
$|\Phi\rangle_A$, then $LL_q^\dag$ is injective and $f_q$ is linear.
We can conclude that the condition that ``the probability of the
measurement outcome~$q$ does not depend on the input state'' (that is,
Alice learns nothing about the input state due to the measurement) is
equivalent to that the teleportation channel is linear.  Moreover, it
can be proven in a way not detailed here that the linearity of the
channel is equivalent to its unitarity---therefore its unitary
reversibility.

\section{Entanglement matching}
\label{sec:matching}

In this section we answer the question what measurements provide fidelity 1
teleportation for an arbitrarily given (even partially) entangled shared
state.  Suppose that the shared state $|\sigma\rangle$ is described by an
invertible antilinear operator $L$.  If Bob applies a unitary transformation
$U_q\colon \HH_C \to \HH_C$ which may depend on the result $q$ of Alice's
measurement, then the final state of system $C$ reads $|\text{out}\rangle_C =
p_q^{-1/2} U_q LL_q^\dag |\Phi\rangle_A$.  Let $i_{AC}$ be a unitary
isomorphism between $\HH_A$ and $\HH_C$ so that we can compare the states of
systems $A$ and $C$.  The teleportation condition is
\begin{equation}
  \frac1{\sqrt{p_q}} U_q LL_q^\dag =  i_{AC}
\end{equation}
which also guarantees that $p_q(|\Phi\rangle_A)$ is independent of
$|\Phi\rangle_A$.  From this we conclude that a measurement with an
outcome described by the antilinear operator
\begin{equation}
  L_q = \sqrt{p_q} i_{AC}^{-1} U_q L^{-1}\strut^\dag
\label{eq:L_q}\end{equation}
supports fidelity 1 conditional teleportation.  The appropriate
recovering unitary transformation applied by Bob is to be $U_q$.

Although $p_q$ in (\ref{eq:p_q}) depends on $L_q$, this can be
resolved by the fact that $L_q$ has a norm ${\mathrm{tr}}_B (L_q^\dag
L_q)=1$.  Then we obtain that the probability is
\begin{equation}
  p_q = \left[ {\mathrm{tr}}_B \left( (L^\dag L)^{-1} \right) \right]^{-1}.
\label{eq:p_q2}\end{equation}

For an arbitrary entangled shared pair described by invertible $L$,
the set of measurement outcomes providing fidelity 1 conditional
teleportation is given by the set
\begin{equation}
  {\mathcal M}_L = \left\{\, L_q =
  \frac{i_{AC}^{-1} U L^{-1}\strut^\dag} 
  {\sqrt{{{\mathrm{tr}}}_B \big( \textstyle L^{-1} L^{-1}\strut^\dag \big)}}
  \, \Bigg| \, \mbox{$U$ is unitary} \,\right\}.
\label{eq:M_L}\end{equation}
Thus not every possible measurement outcome
allows teleportation, only those described by ${\mathcal M}_L$.  The
measurement and the shared state should be ``matched'' to each other.
This can be regarded as a generalization of ``entanglement matching''
introduced in Ref.~\cite{pra61_034301}.

It is worth to note that (\ref{eq:p_q2}) is the same for every outcome $q$
that matches the shared state in the above sense.  The probability of a
successful outcome depends only on the shared state.  Another important result
is that the set ${\mathcal M}_L$ of matching outcomes is spanned by local
unitary transformations: if one finds a measurement outcome which enables
probabilistic teleportation, then every matching outcome can be obtained from
it by a local unitary transformation on system $A$.

We give an example for entanglement matching.  Suppose that the
antilinear operator $L$ describing the shared state
$|\sigma\rangle_{BC}$ is given by the following matrix:
\begin{equation}
  L = \left( \begin{matrix} \alpha_1\cr & \ddots\cr &&\alpha_n\end{matrix} \right),
  \qquad |\sigma\rangle_{AB}=\sum_i \alpha_i | i \rangle_B | i \rangle_C,
\end{equation}
where all $\alpha_i$ are nonzero (consider a Schmidt decomposition for
example).  Taking that the unitary transformation $U_q$ is identity,
we obtain from (\ref{eq:L_q}) that a matching measurement outcome is
given by
\begin{eqnarray}
  L_1&=& \left( \sum_i \frac1{|\alpha_i|^2} \right)^{-1/2} \left(
  \begin{matrix}1/\alpha_1^\ast \cr & \ddots\cr && 1/\alpha_n^\ast\end{matrix} \right),
  \nonumber\\
  |\sigma_1\rangle_{AB} &=& \left( \sum_i \frac1{|\alpha_i|^2}
  \right)^{-1/2}  \sum_i \frac1{\alpha_i^\ast}  | i \rangle_A | i \rangle_B.
\end{eqnarray}
Thus if Alice measures an observable with an eigenstate equal to
$|\sigma_1\rangle_{AB}$ then that measurement outcome implements a
conditional teleportation.

\section{Conclusion}
\label{sec:concl}

We have summarized the relations between quantum channels, bipartite states
and antilinear operators, focusing the description of bipartite pure states
with the latter. Applying this description, we have characterized all possible
conditional teleportation schemes. We have found that the independence of the
probability of a measurement outcome on the input state is a necessary and
sufficient condition of the \emph{linearity} of the transformation to be
applied by the receiver. We have generalized the concept of ``entanglement
matching'', which means that in schemes under consideration the entangled
state shared by the parties, and those measured by the sender should ``match''
each other.

The results presented here show that this formalism is applicable of
describing entanglement and quantum teleportation in a quite general way. This
method may be also applicable for the treatment of entanglement between
systems described by Hilbert-spaces of different dimensionality. It can also
have consequences regarding the description of teleportation and related
phenomena in the framework of quantum operations.

\begin{acknowledgements}
This work was supported by the Research Fund of Hungary (OTKA) under contract
No. T034484. One of the authors (M. K.) thanks Prof. Vladim\' \i r Bu\v zek
for useful discussions.
\end{acknowledgements}

\end{document}